\documentclass[journal]{IEEEtran}
\ifCLASSINFOpdf
\else
\fi
\usepackage{tabularx}
\usepackage{makecell}
\usepackage{graphicx}
\usepackage{subfig}
\usepackage{caption}
\usepackage[font=footnotesize]{caption}
\usepackage{amsmath}
\usepackage[switch,pagewise]{lineno} 
\usepackage{lipsum} 
\usepackage[flushleft]{threeparttable}
\usepackage{xcolor}
\usepackage{soul}
\usepackage{multirow}
\usepackage{alterqcm}
\usepackage{array}
\usepackage{comment}

\begin{document}
%
\title{Investigating HMIs to Foster Communications between Conventional Vehicles and Autonomous Vehicles in Intersections}
%
%
%

\author{{ Lilit Avetisyan, Aditya Deshmukh, X. Jessie Yang, Feng Zhou}

\thanks{L. Avetisyan, A. Deshmukh and F. Zhou are with the Department of Industrial and Manufacturing, Systems Engineering, The University of Michigan-Dearborn, Dearborn, 4901 Evergreen Rd. Dearborn, MI 48128 USA (e-mail: \{lilita, adityd, fezhou\}@umich.edu). X. Jessie Yang is with the Departmetn of Industrial and Operations Engineering, University of Michigan, Ann Arbor (email: xijyang@umich.edu)}
\thanks{Manuscript received May 28, 2023; revised xxx xx, 2023.}}

%
%

\markboth{IEEE Transactions on Human-Machine Systems,~Vol.~xx, No.~xx, Month~2023}%
{Avetisyan \MakeLowercase{\textit{et al.}}: Investigating HMIs to Foster Communications between Conventional Vehicles and Autonomous Vehicles in Intersections}
%

\maketitle

\begin{abstract}
In mixed traffic environments that involve conventional vehicles (CVs) and autonomous vehicles (AVs), it is crucial for CV drivers to maintain an appropriate level of situation awareness to ensure safe and efficient interactions with AVs. This study investigates how AV communication through human-machine interfaces (HMIs) affects CV drivers' situation awareness (SA) in mixed traffic environments, especially at intersections. Initially, we designed eight HMI concepts through a human-centered design process. The two highest-rated concepts were selected for implementation as external and internal HMIs (eHMIs and iHMIs). Subsequently, we designed a within-subjects experiment with three conditions, a control condition without any communication HMI, and two treatment conditions utilizing eHMIs and iHMIs as communication means.  We investigated the effects of these conditions on 50 participants acting as CV drivers in a virtual environment (VR) driving simulator. Self-reported assessments and eye-tracking measures were employed to evaluate participants' situation awareness, trust, acceptance, and mental workload. Results indicated that the iHMI condition resulted in superior SA among participants and improved trust in AV compared to the control and eHMI conditions. Additionally, iHMI led to a comparatively lower increase in mental workload compared to the other two conditions. Our study contributes to the development of effective AV-CV communications and has the potential to inform the design of future AV systems.

\end{abstract}


%
\IEEEpeerreviewmaketitle

\section{Introduction}

In recent years, the potential of autonomous vehicles (AVs) has been extensively discussed, as they hold the promise of revolutionizing transportation by reducing accidents caused by human error, enhancing transportation efficiency, and increasing mobility for individuals who are unable to drive \cite{ayoub_from_2019, Zhou2020_IJHCI}. However, a significant challenge in fully realizing the benefits of AVs lies in establishing effective communication between AVs and other road users \cite{avetisyan2022SA, Jayaraman2018_HRI, Jayaraman2019_FRA}.

In response to the challenge, many studies have focused on the interactions between AVs and vulnerable road users, including pedestrians, bicyclists, and motorcyclists, with an emphasis on designing external human-machine interfaces (eHMIs) through the use of both visual (e.g., lights and laser projections) and auditory cues (e.g., low frequency hum, chime, beep) \cite{bai2021investigating, palmeiro2018interaction, rasouli2019autonomous}. The eHMIs could help the AV to convey its intentions to vulnerable road users in complex situations to reduce misunderstandings and improve traffic safety. For example, Bai et al. \cite{bai2021investigating} explored different external interaction modalities, including visual, auditory, and visual+auditory to promote effective communications between AVs and vulnerable road users. Unlike implicit signals through the trajectory and speed of AVs \cite{fuest2018using}, eHMIs adopt a communication strategy to share its intention explicitly, which improves the effectiveness of pedestrians' crossing decisions and cuts down on the amount of time it takes to cross the street \cite{dietrich2019projection}. Hensch et al. \cite{hensch2020ehmiComm} investigated the perceived safety of eHMIs and pointed out that it increased the perceived trustworthiness and intuitiveness of AVs at the same time.

In addition to investigating the interactions between AVs and vulnerable road users, it is also crucial to examine the interactions between AVs and CVs. As AVs will coexist with CVs on the roads for the foreseeable future, effective communication between these vehicle types is essential for ensuring smooth and safe traffic flow. For instance, when an AV approaches a four-way intersection, it must communicate with other vehicles to ensure their awareness of its presence and facilitate appropriate speed and position adjustments \cite{papakostopoulos2021effect}. 



However, in contrast to the extensive research examining interactions between AVs and vulnerable road users, there is little research examining  communication between AVs and CVs, with only few exceptions. This is disturbing as traditional means of communication between human drivers, such as waving or gestures, will highly like become  ineffective in mixed traffic environments \cite{juhlin1999traffic}, as the passenger in an AV may not be available to engage in such communication. 

In this study, we aimed to fill the research gap and investigate whether human drivers in CVs could benefit from AVs equipped with appropriately designed HMIs in mixed traffic, especially in challenging traffic situations. We designed a within-subjects experiment in VR where the treatment was the HMIs with three different designs, including a control condition without any HMI, an eHMI located on the front of the AV, denoted as eHMI, and an HMI located in the internal of the CV, denoted as iHMI. We designed two scenarios where the design showed the AV would yield its right of way or insist on its right of way to show how these three design variations impacted the participants' situation awareness (SA) of road conditions, their trust in AVs, the acceptance of such designs, and mental workload involved in the negotiation process using both self-reported measures and eye tracking data.

\section{Related Work}
\subsection{Communications between AVs and vulnerable road users} 

Due to the increase in AVs on the road, it is necessary to develop a communication system between AV and other road users, especially vulnerable road users, such as pedestrians and bicyclists, to increase road traffic efficiency and safety. There are certain situations, such as four-way intersections \cite{papakostopoulos2021effect}  and pedestrian crosswalks \cite{fuest2018using}, in which the need for communication between AVs and other road users increases. There are several different ways that AVs could communicate with vulnerable road users, including implicit and explicit communications. Implicit communications make use of the AV's ability to detect and respond to the presence of other road users without direct communication, but rather using existing implicit cues, such as vehicle trajectory and speed changes \cite{miller2022implicit}. For example, Dietrich et al. \cite{dietrich2019implicit} examined the effects of pitch and deceleration of AV on pedestrian crossing behavior in urban scenarios and showed that changes in pitch and deceleration of AV had a significant impact on pedestrian crossing behavior, with higher pitch and deceleration leading to more cautious crossing behavior by pedestrians. While such implicit communications seem dominant in modern driving, it might be not adequate in ambiguous situations between AVs and vulnerable road users \cite{tabone2021vulnerable}. 

Another important communication strategy is to explicitly display the intention on an eHMI so that other road users can understand AV's intention easily. These eHMIs come with different modalities, including visual cues, such as LED strips and screens, laser projections on the road, and auditory signals, such as beeps and chimes. For example, Bai et al. \cite{bai2021investigating} examined six different visual cues and five different auditory cues and found that participants preferred pedestrian silhouette on the front of the AV and chime to other cues and a combination of both visual and auditory cues were preferred the most. 

While numerous studies demonstrate that eHMIs improve the interaction between pedestrians and vehicles, it is still unclear exactly what information an eHMI should show. Merat et. al. \cite{merat2018externally} investigated the information needs of vulnerable road users and found that messages about vehicle action, e.g, turning, starting, stopping,  were more important than vehicle dynamics, such as speed. Additionally, they pointed out that auditory- and light-based communication methods were preferable than text-based messages. Schieben et al. \cite{schieben2019designing} examined the design of the interaction of AVs with other traffic participants, focusing on human needs and expectations. The study emphasized the importance of effective communication and understanding between AVs and other road users, and the need for AVs to adapt to the behavior and expectations of vulnerable road users. Faas et al. \cite{faas2020external} investigated the information that should be displayed on the eHMI of AVs and showed that participants believed that information on the vehicle's speed and direction, as well as information on the vehicle's state, such as whether it was in autonomous mode or not, should be prominently displayed. 


The design of eHMI should also take into account factors such as trust, acceptance, and user experience to improve the overall interaction between AVs and other road users by considering human needs and expectations in the design of AVs and their interactions with other road users, especially the vulnerable ones.  Eisma et al. \cite{eisma2021external} conducted an experiment about different message perspectives (i.e., egocentric and allocentric) displayed on eHMIs and examined their effects on participants' understanding. They found that pedestrian-centric perspective was more effective in communicating the vehicle's intent and was better understood by the participants. 
Lee et al. \cite{lee2021negative} found that certain eHMI designs could lead to confusion and mistrust among pedestrians, resulting in them taking longer to cross the street and making more mistakes due to the fact that these designs confused the participants. Therefore, it is important that design of  eHMIs for AVs should be carefully considered to ensure that they effectively communicate the vehicle's intent and do not negatively impact other road users' behavior.

\subsection{Communications between AVs and CVs} 
As AVs become more prevalent, they will inevitably share the road with manually driven vehicles or CVs. This mixed traffic environment presents a key challenge of communicating between these two types of vehicles and they must be addressed to ensure the safety and efficiency of the transportation system. However, there is a lack of research in investigating the communication between AVs and CVs, especially in complex and ambiguous situations, such as intersection without traffic lights, bottleneck scenarios, where traditional vehicle communications often fail. Aoki and Rajkumar \cite{aoki2018dynamic} investigated dynamic intersection scenarios with possible collision risks and showed that using V2V communications and sensor systems, AVs effectively communicated with other road users and positively impacted on traffic flow. Rettenmair et al. \cite{rettenmaier2020after} studied different strategies of communication between AVs and CVs in a bottleneck scenario where two cars came facing each other while only one was able to pass. The results showed that participants indicated better visibility of the display in comparison with the projection. Clearly visible displays were better instead of laser projections, which were not easily legible at greater distances \cite{eisma2021external}. In another study, Rettenmaier et al. \cite{rettenmaier2019passing} found that a well-designed eHMI using two arrows to indicate the direction had a positive impact on comprehensibility, transferability and simplicity and the participants had significantly shorter passing times and fewer crashes compared to other eHMI designs. Papakostopoulos et al. \cite{papakostopoulos2021effect} investigated the effect of eHMIs on drivers' ability to infer the motion intention of AVs. They found that the presence of eHMIs on the AV accelerated CV drivers’ decision-making in terms of improving participants’ driving behavior and reducing the overall crossing time. Therefore, eHMIs can have a positive impact on other vehicles' behavior by carefully considering the design of eHMIs for AVs to ensure that they effectively communicate the AV's intention and not negatively impact the safety of other road users.



\section{Method}

\subsection{Designing HMIs with a human-centered approach}
In the first phase of this study, we employed a human-centered design process \cite{ayoub2020otto} to develop interfaces to facilitate communication between CVs and AVs for a specific scenario in intersections. To empathize with CV drivers and understand their needs, we reviewed existing literature and had a class discussion (with more than 30 master students in the human-centered design program in University of Michigan-Dearborn) about the challenges faced by communications between CVs and AVs. Based on our analysis, we identified three main challenges, including: 1) conspicuousness - easy to see visually, 2) comprehensibility - easily understandable with minimal cognitive effort, and 3) identifiability - easy to recognize to whom it was addressed. 

\begin{figure*}[bt!]
\centering
\includegraphics[scale = 0.6]{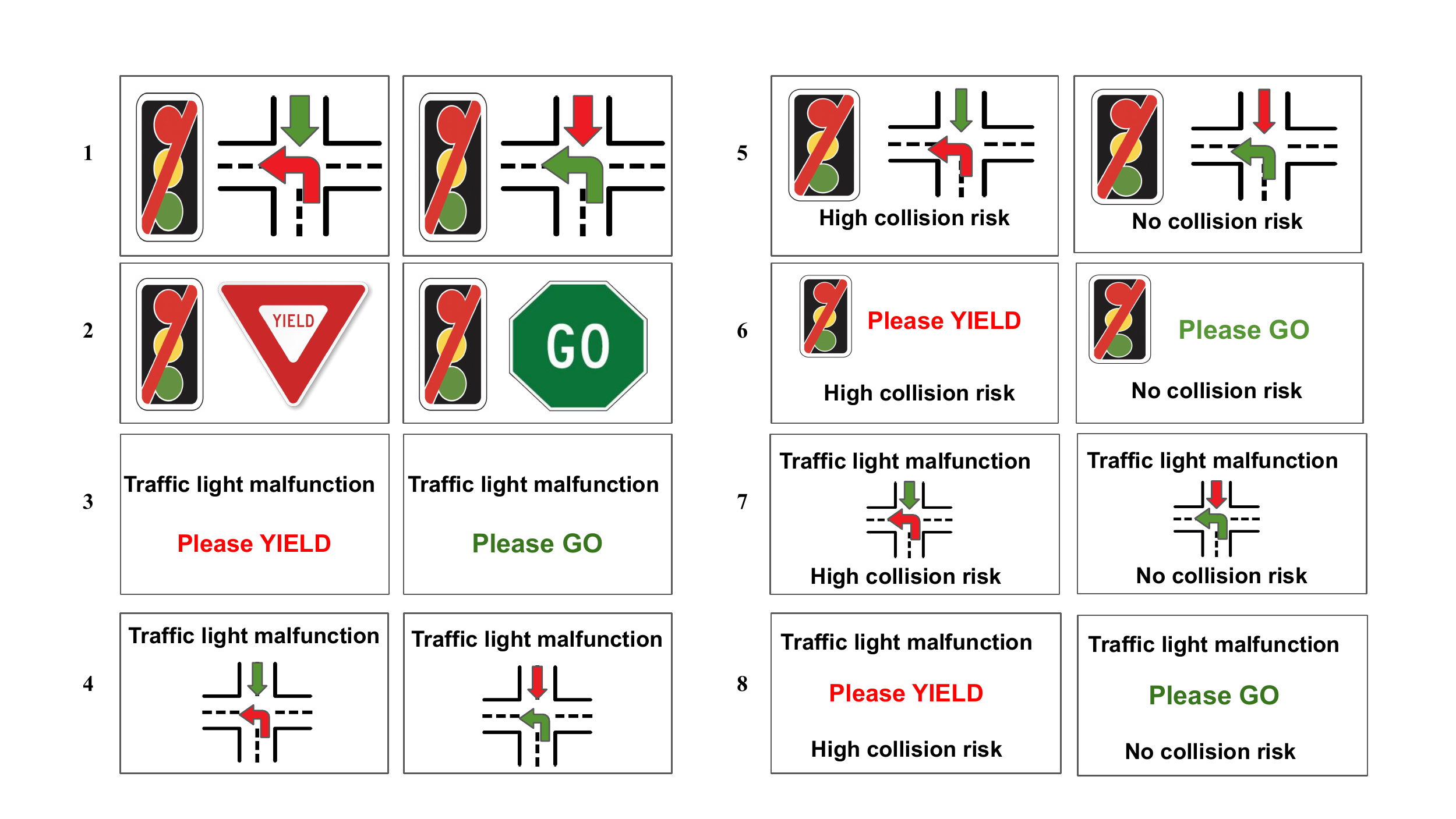}
\caption{Interface design concepts for AV-CV communication for a left turn scenario in an intersection, where the traffic lights were malfunctioning. In all the concepts, the left panel in each design shows that the CV yields the right of way to the AV, while the right panel shows that CV insists the right of way.} 
\label{fig:concepts}
\end{figure*}

Then, we had a brainstorm session to generate a wide range of possible ideas for the ``left turn" scenario in an intersection and followed the design principles outlined in Rettenmaier et al.'s study \cite{rettenmaier2020after} and Avetisyan et al.'s SA framework \cite{avetisyan2022SA} for informational needs. Ultimately, we narrowed down to eight design concepts (see Fig. \ref{fig:concepts}) that included two versions for the AV to 1) yield and 2) insist on the right of way, and were presented using different visual formats, i.e., signs, texts, or a combination of both. Designs 1 to 4 included SA level 1 (i.e., perception of the items in the environment) and level 2 (i.e., comprehension of the current situation) information, while designs 5 to 8 included additional SA level 3 information (i.e., projection of future status of the environment). To facilitate communication between CVs and AVs in the selected scenario, we broke down the HMI message into three parts, which explained the traffic situation, a suggested way of behaving, and improved trust. Firstly, we added a sign or text that described the current issue that the traffic lights were malfunctioning. Secondly, we included right-of-way information that informed CV drivers how to proceed, incorporating well-known traffic signs and colors to enhance comprehension. Thirdly, we attempted to provide extra information to increase the confidence of CV drivers.

To evaluate the interface designs, we conducted an online survey study with 32 participants who were introduced to the ``left turn" scenario and asked to show their level of agreement for the prototypes of the design concepts using a 7-point Likert scale based on four statements: 1) The message is easy to understand, 2) The message contains relevant details, 3) The message helps to respond quickly, and 4) The message is preferred with text only. The final score was calculated as the average of these four statements. Among the eight designs, the second interface design (see Fig. \ref{fig:concepts}) received the highest rating of 5.28 and was chosen to be tested in the second phase of the study using a driving simulator in a VR environment.

\subsection{Participants}
In the design phase, a total number of 32 students were participated and evaluated eight concepts. At this phase, participants were compensated with 1\% course credit.

In the experiment phase, a total number of 50 participants took part in the experiment. Due to the severe motion sickness and eye calibration failures, six participants did not finish the experiment. Therefore, the data analysis was conducted based on the remaining 44 participants (11 females and 33 males; Age = 24.4 ± 4.19 years old) who were university students located in the United States and possessed a valid U.S. driver's license. On average, participants had 5 ± 4.52 years of driving experience and drove approximately 5 ± 1 days per week. This study was conducted in accordance with the ethical requirements of the Institutional Review Board at the University of Michigan (application number HUM00219554). Participants received a compensation of \$20 in cash upon completion of the study, and the average completion time was 35 minutes for male and 41 minutes for female participants. Participants experienced low levels of motion sickness in the experiment (i.e., mean ratings were 1.66 and 1.70 for male and female participants on 7-point Likert scales).

\subsection{Apparatus} 
The experiment was conducted in a virtual reality driving simulator at the University of Michigan-Dearborn using a desktop computer with an Intel Xeon(R) W-2104 CPU  processor running at 3.20GHz, 64.0 GB of RAM, and an NVIDIA GeForce RTX 3060 graphics card with 12 GB of memory. The operating system used was Windows 10. The experiment employed an HTC Vive Pro Eye headset (Taipei, Taiwan) in combination with a Logitech G29 Driving Force (Lausanne, Switzerland) steering wheel and floor pedal. Four different drives (one for training and three for experimentation) were created using the Unity game engine (San Francisco, CA). All self-reported data was collected using a survey developed and administered through the Qualtrics (Seattle, WA) platform.


\subsection{Experiment Design}

\textbf{Independent variables.} In this study, a within-subjects experiment was conducted, in which the independent variable was the communication interface with three conditions: control, eHMI, and iHMI. The control condition did not have any explicit communication between the AV and the CV. In the eHMI condition, the AV communicated with the CV through an external display attached to the front of the AV, which shared the current traffic situation and right-of-way information from the egocentric perspective (the CV driver's perspective). In the iHMI condition, the AV communicated with the CV through an internal display that appeared on the CV's windshield and shared the same information as described for eHMI.

\begin{figure}[bt!]
\centering
\includegraphics[width=0.8\linewidth]{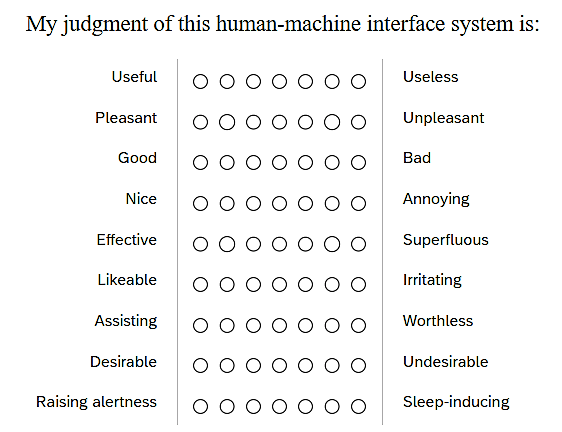}\hfill
\caption{The acceptance scale \cite{van1997acceptance}. The usefulness measure was the average of useful, good, effective, assisting, and raising alertness items. The satisfaction measure was the average of pleasant, nice, likable, and desirable items.}
\label{fig:acc_scale}
\end{figure}

\textbf{Dependent variables.} The study collected both self-reported and eye tracking data. Self-reported measures were used to assess SA, trust in AV, interface acceptance. \textit{Situation awareness} was assessed using the Situation Awareness Global Assessment Technique (SAGAT) technique \cite{endsley1988design}, where the SA questions were designed following the method used in previous studies \cite{avetisyan2022SA, vandebBeukel2017SA_assess}. After each trial, the participants were asked to answer SA questions (see Table \ref{table:SA}) for two interactions separately, where they chose all the applicable answers out of the 5 possible options. The final SA score was measured based on the number of correct answers, with a score range of 0 to 3. To prevent negative impacts on participants' engagement in VR and to avoid causing additional motion sickness, the SA was evaluated in the post-trial phase. \textit{Trust} in AV was measured with Jayaraman et al.'s \cite{jayaraman2019trustscale} trust scale with 7-point Likert scales. At the end of each session, participants evaluated their trust in five dimensions: competence, predictability, dependability, responsibility, reliability over time and faith, by answering question presented in Table \ref{table:trust}. To understand the acceptance of the proposed concepts as communication interfaces, Van der Laan et al.'s \cite{van1997acceptance} nine-item acceptance measure was applied, where participants rated them in two dimensions: 1) perceived usefulness (i.e., useful, good, effective, assisting, and raising alertness), which focused on the functional aspects of the concepts and how well it could assist the participant in current situations, and 2) satisfaction (i.e., pleasant, nice, likable, and desirable) which focused on the overall emotional responses and fulfillment of expectations after experiencing the concepts. The measures were assessed by 7-point Likert scales (see Fig. \ref{fig:acc_scale}) after each trial was completed. Furthermore, the participants' simulation sickness was evaluated using the Simulator Sickness Questionnaire (SSQ) \cite{kennedy1993ssq} on a 7-point Likert scale.

To collect eye-tracking data from participants, the integrated eye tracker in the HTC VIVE Pro Eye headset was used with a resolution of 1440 x 1600 pixels per eye, and a 110-degree horizontal field of view. The eye tracker had a 120 Hz frequency of gaze data output with an accuracy of 0.5-1.1 degrees. In total, 7 measures were collected and used to analyze the pupil diameter and eye openness (see Table \ref{table:eyeMeasures}).
Previous studies showed that mean change in pupil diameter was a reliable measure since it can eliminate the side factors, i.e. environment illumination, that could potentially influence on pupil diameters \cite{palinko2012pupilchange}. 

Therefore, to investigate how participants’ mental workload and vigilance \cite{chen2014pupilMW, martin2022pupillometry} were influenced by the interface conditions, mean pupil diameter and eye openness changes were examined. Prior to the analysis, the raw data went through four step of data processing. First, the invalid eye data rows and outliers were removed. Next, the records of left and right eyes were combined by their mean coordinate values. Third, the interaction moments were identified using vehicle positions in the VR and the data was segmented to ten second before and after the interaction with AV. Finally, the mean pupil diameter change and eye openness change were calculated per interaction, and the average of two interactions was used for further analysis.


\begin{table}[ht]
\caption{An example question for the iHMI condition to measure SA with a SAGAT questionnaire \cite{endsley1995sagat}, where the 1), 3), and 4) options were the correct answers. }
\small
\centering

\begin{tabular}{m{0.45\textwidth}}
\hline\hline

Question\\
\hline
\vspace{.5ex}
\begin{center}

\includegraphics[width=.8\linewidth]{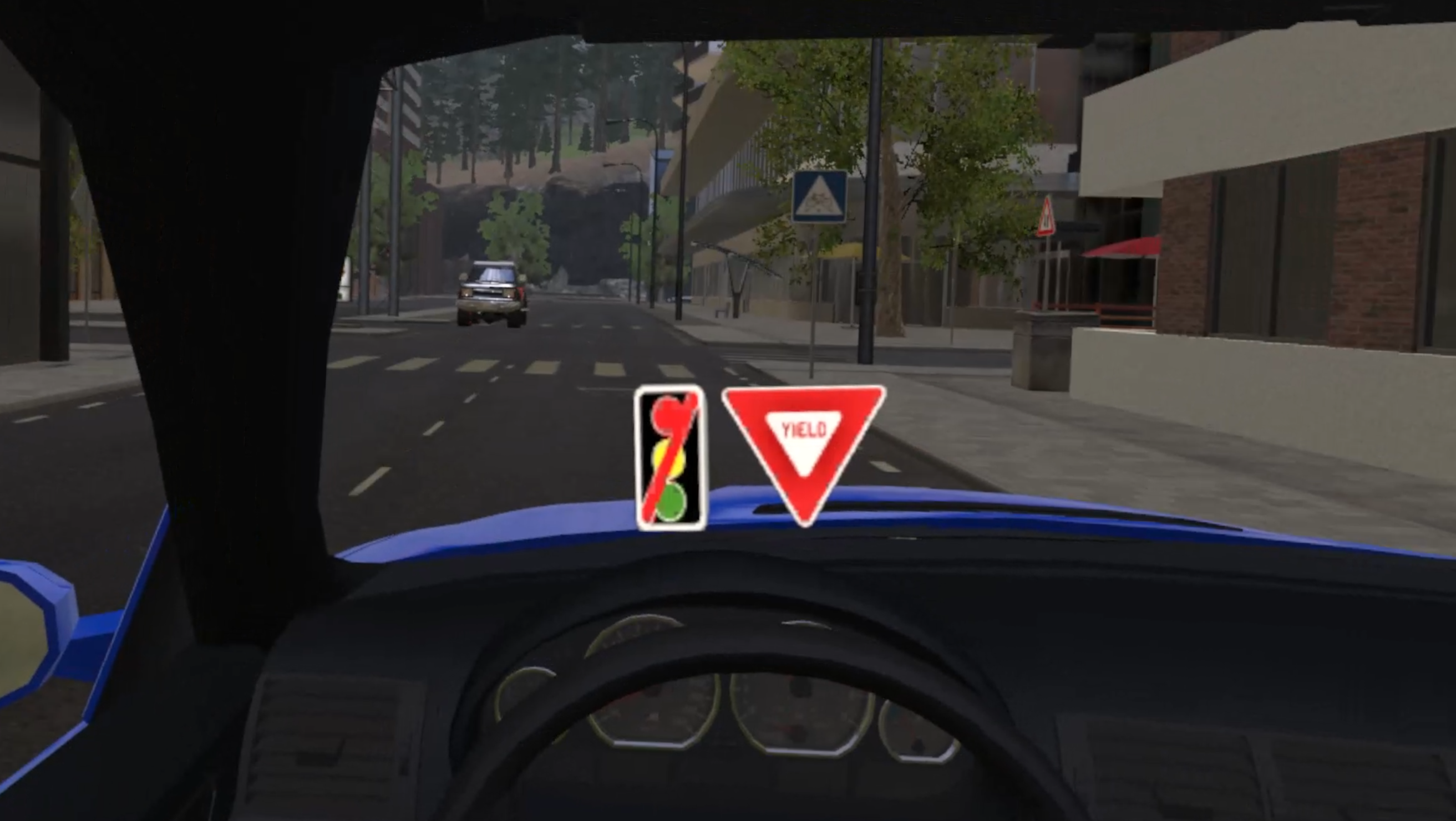} 
\end{center}
You and the incoming automated vehicle approached an intersection as shown in the picture. Based on your understanding, what was the situation at the intersection (pick all applicable)?
\vspace{.5ex}
 \begin{enumerate}
 \item The traffic lights are malfunctioning
 \item The incoming automated vehicle will yield 
 \item The incoming automated vehicle has the right of way
 \item The incoming automated vehicle expects you to yield
 \item The incoming automated vehicle violated traffic rules
\end{enumerate} 
\\ \hline \hline
\end{tabular}

\label{table:SA}

\end{table}




\begin{table}[h]
\caption{The trust questionnaire \cite{jayaraman2019trustscale}.}
\small
\centering
\begin{tabular}{p{0.2\linewidth}p{0.7\linewidth}}
\hline\hline

 Dimension & Question \\ 
 \hline
 
Competence & To what extent did the autonomous cars perform their function properly i.e., recognizing you and reacting for you?\\

Predictability & To what extent were you able to predict the behavior of the autonomous cars from moment to moment?\\

Dependability & To what extent can you count on the autonomous cars to do its job?\\

Responsibility & To what extent the autonomous cars seemed to be wary of their surroundings?\\

Reliability over time & To what extent do you think the autonomous car's actions were consistent through out the interaction?\\

Faith & What degree of faith do you have that the autonomous cars will be able to cope with all uncertain ties in the future?\\

\hline
\hline
\end{tabular}
\\[1ex]
\label{table:trust}
\end{table}

\begin{table}[h]
\caption{Eye tracking measures. Each measure was collected for the left and right eye separately.}
\small
\centering
\begin{tabular}{p{0.25\linewidth}p{0.7\linewidth}}
\hline\hline

 Measures & Description \\ 
 \hline
Timestamp & The current time of data recording\\
Eye validity & The variable that explains validity of eye data\\
Eye openness & The level of eye openness\\
Pupil diameter & The diameter of pupil \\
\hline
\hline
\end{tabular}
\\[1ex]
\label{table:eyeMeasures}
\end{table}

\begin{figure}[ht!]
\centering
\includegraphics[width = 0.9\linewidth]{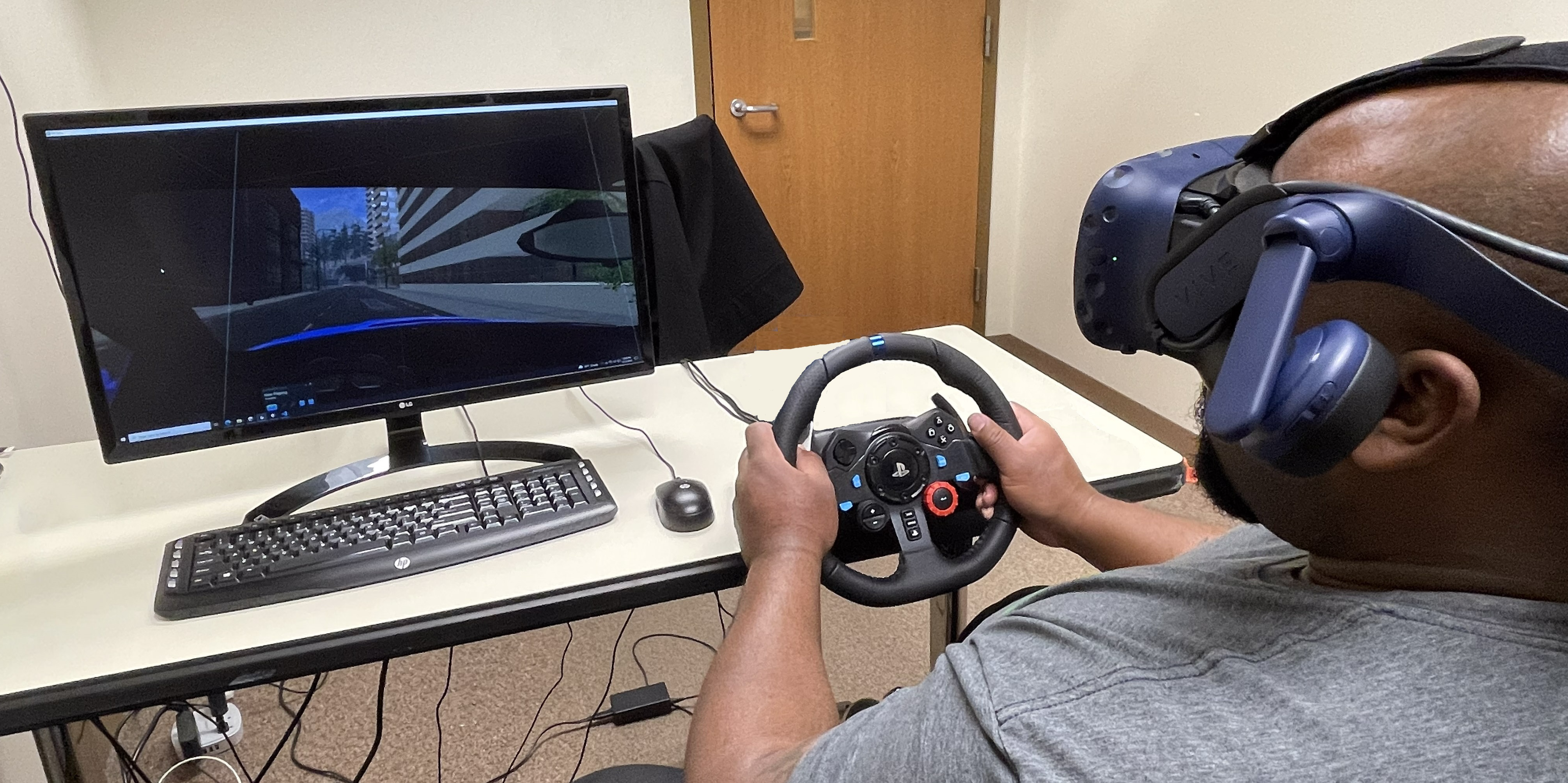}
\caption{Experiment setup with VR driving simulator. The participant was wearing HTC Vive headset and driving a CV via Logitech G29 steering wheel and pedals. The screen showed the participant's view in the VR environment.}
\label{fig:expSetup}
\end{figure}

\subsection{Survey Design and Procedure}
The experiment consisted of six sections. In the first section, participants completed the demographic section of the survey and received an introduction to the experimental process and tools. Following the introduction, the experimenter calibrated the eye tracker, and the participants had a training session where they had an opportunity to familiarize themselves with the VR environment and driving equipment through a test drive (see Fig. \ref{fig:expSetup}). During this training session, participants experienced interactions with AVs similar to the experimental scenarios and were introduced to two interface concepts. Upon completion of the training session, participants were given the choice to either continue or stop their participation in the experiment.
Since a within-subjects experimental design was employed, after the training sessions, participants experienced each of the three experiment sessions corresponding to three conditions, i.e., eHMI, iHMI, and control conditions, in a randomized order. At the beginning of each session, the participants received brief introduction about current interface. During the drive, the eye tracking outputs and vehicle trajectory were recorded. At the end of each session, the simulation was paused and the participants filled in the survey section required to measure the dependant variables (i.e.,  SA, trust, and acceptance).
At the end of the last session, participants were asked to evaluate the 16 symptoms of simulation sickness listed on the SSQ.
Overall, the experiment took approximately 38 minutes to complete, with 15 minutes allotted for demographic information, instruction, and training, and approximately 5 minutes for each experimental session. The participants also had the option to take a 5-minute break between each experiment condition.

\subsection{Scenario Design}
\begin{figure}[h]
\centering
\includegraphics[scale=0.6]{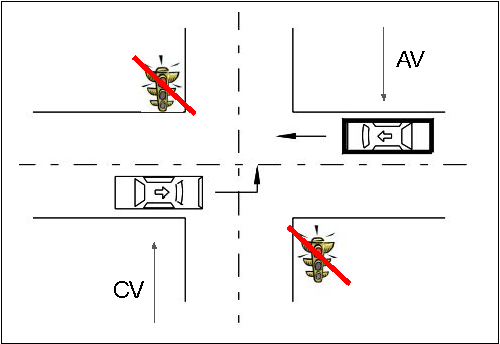}
\caption{The CV is turning left at an intersection with malfunctioning traffic lights and has to turn left, crossing the path of the incoming AV.  The scenario was tested in two different AV behaviors. In one scenario, the AV yielded to the CV, while, in the other, the AV insisted on the right of way.}
\label{fig:scenario}
\end{figure}

To develop an interaction scenario, a high-risk traffic situations was investigated that resulted in communication issues or even crashes. A typical scenario was selected from Najm et al.'s crash report \cite{najm2007crashscenarios}, which analyzed police-reported crashes and categorized them based on five metrics: frequency of occurrence, number of people involved, injury severity, economic cost, and functional years lost. The scenarios was adapted for our experiment to focus on the communication between AVs and CVs. 
In each scenario, there were two variations with an AV at two different intersections where the CV driver was instructed to turn left by the integrated navigation system's voice (as shown in Fig. \ref{fig:scenario}). In one interaction, the AV yielded its right of way, while in the other interaction, it insisted on the right of way. When the distance between the AV and the CV reached the predefined distance, the AV displayed a message informing that the traffic lights were not functioning, and the CV either had the right of way or had to yield to the AV.

\begin{figure} [bt!]
	\centering
	\subfloat[\label{fig:i-go}]{\includegraphics[width=.45\linewidth, height=.3\linewidth]{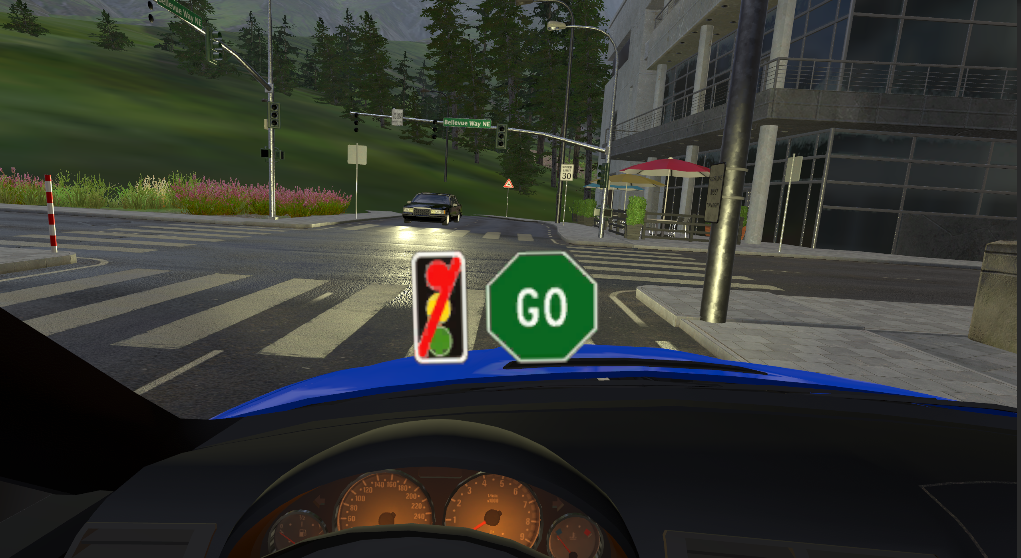}}
	\hspace{0pt}
	\subfloat[\label{fig:i-yield}]{\includegraphics[width=.45\linewidth,height=.3\linewidth]{figures/iHMI-yield.png}}\\
        \subfloat[\label{fig:e-go}]{\includegraphics[width=.45\linewidth,height=.3\linewidth]{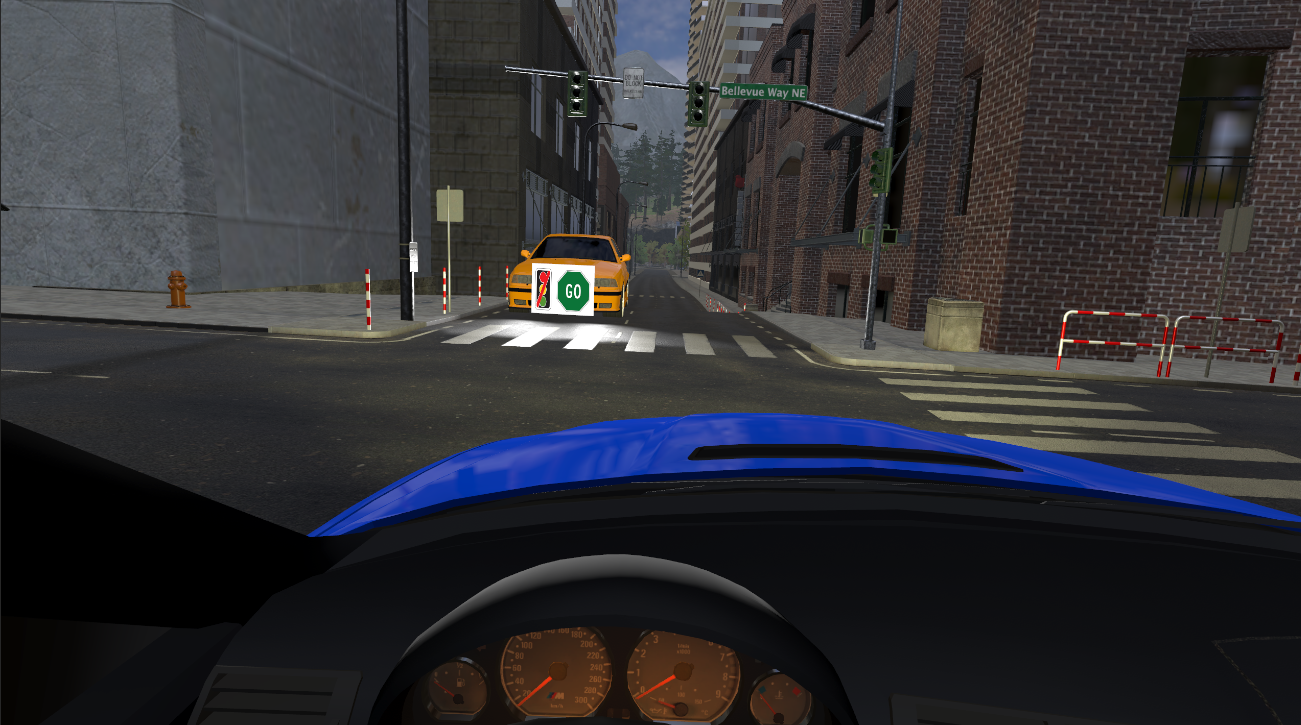}}
	\hspace{0pt}
	\subfloat[\label{fig:e-yield}]{\includegraphics[width=.45\linewidth,height=.3\linewidth]{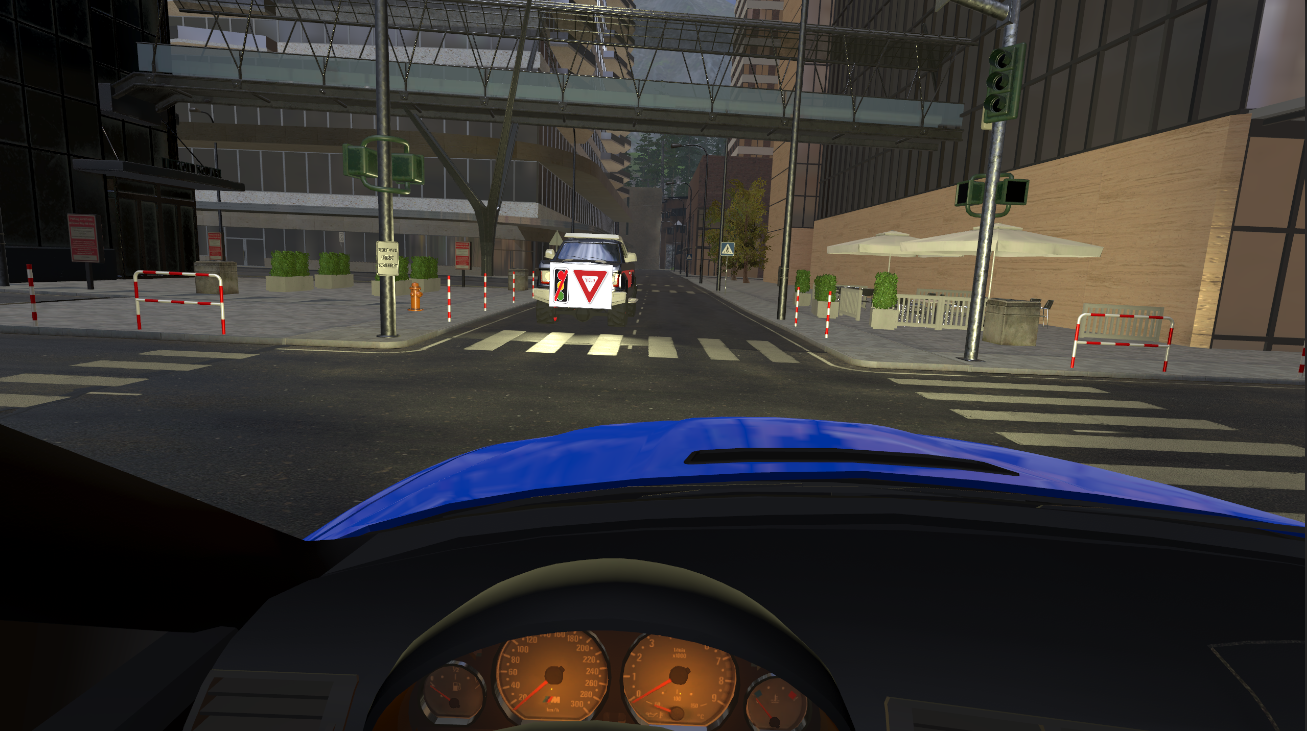}}
    \caption{Experimental scenarios with (a) go and (b) yield messages in iHMI condition, (c) and (d) in eHMI condition.}\label{fig:drivingScenarios}
\end{figure}

In the eHMI condition, the messages were displayed on the front of the AVs as shown in Fig. \ref{fig:e-go} and \ref{fig:e-yield}, disappeared as the vehicles passed each other. In the iHMI condition, the message was displayed on the CV's windshield as shown in Fig. \ref{fig:i-go} and \ref{fig:i-yield}. In the control scenario, the AV did not have a communication interface, and the CV driver had to rely on their own understanding to navigate.

\section{RESULTS} 
As the experiment was conducted using a within-subjects experimental design, a repeated measures one-way analysis of variance (ANOVA) was conducted to examine how the interface conditions affected the dependent variables. For the post-hoc analysis, a Bonferroni correction method was applied. The statistical analyses were carried out using MATLAB and R programming languages, with an alpha level of .05 set for all the tests. The effects of the interface conditions on the dependent variables were assessed using $\eta^{2}$, i.e., the proportion of variance in the dependent variables.

\subsection{Situation awareness}
\begin{figure}[ht]
\centering
\includegraphics[width = 0.9\linewidth]{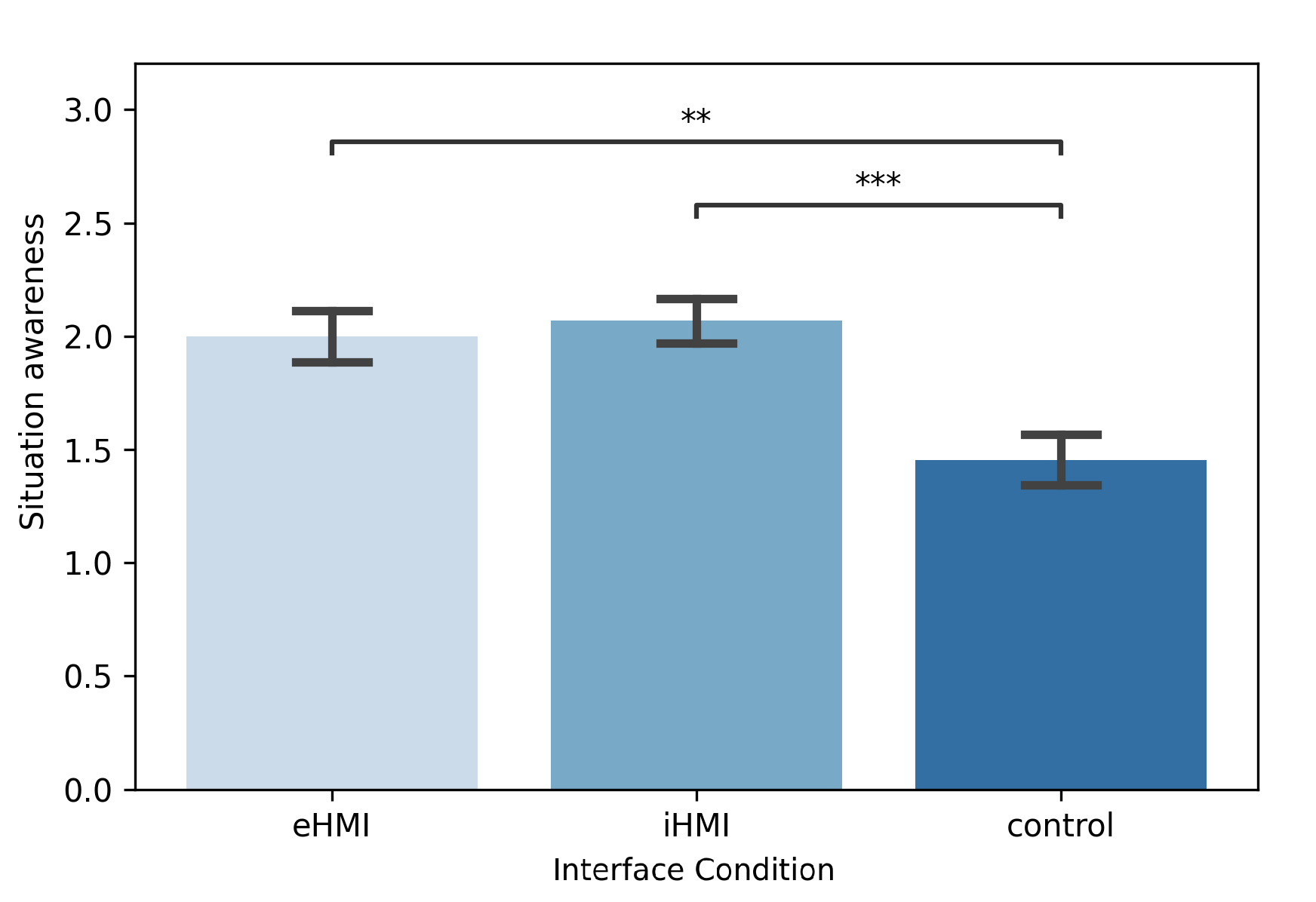}
\caption{The effect of interface conditions on the SA level of the driver in the CV, where ‘**’ indicates $p < .01$ and ‘***’ indicates $p < .001$. Note that the error bar showed the standard error.}
\label{fig:sa}
\end{figure}

SA was measured at two interactions for each interface condition and the mean SA score was used in the analysis. The results of one-way repeated measures ANOVA showed that there was a significant difference among three conditions ($F(2,131) = 13.64, p = .000, \eta^{2} = .132$). As illustrated in Fig.  \ref{fig:sa}, the post-hoc analysis indicated that sharing the information through eHMI ($p=.002$) and iHMI ($p=.000$) significantly increased the SA level of the driver in the CV compared to the control condition. The difference between iHMI and eMHI was not significant.

\subsection{Trust}

\begin{figure}[ht]
\centering
\includegraphics[width = 0.9\linewidth]{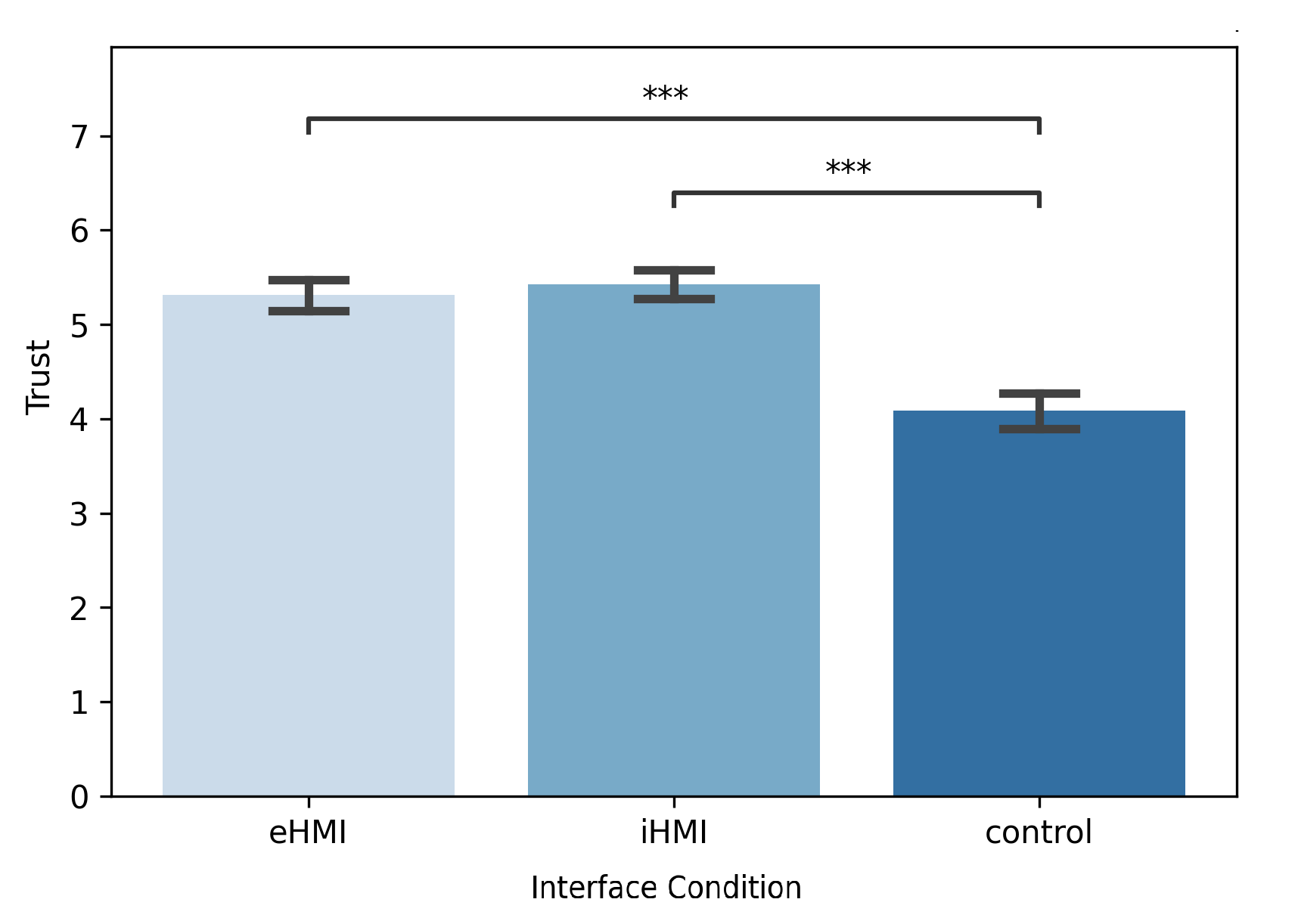}
\caption{The effect of interface conditions on trust in AVs, where ‘***’ indicates $p < .001$. Note that the error bar showed the standard error.}
\label{fig:trust}
\end{figure}

In order to understand the effects of three interface conditions on trust, the mean of the five trust dimensions was analyzed.  The results indicated that trust was significantly different among the tested conditions ($F(2,131) = 25.20, p = .000, \eta^{2}= .233$). Specifically, the trust level was significantly lower in the control condition than that in the eHMI and iHMI conditions ($p = .000$) as shown in Fig. \ref{fig:trust}. Also, in the eHMI condition the reported trust level was lower than iHMI,  but the difference was not statistically significant  ($p=.648$).

\subsection{Acceptance}
Regarding the acceptance of designed interface concepts, the results showed that participants' ratings were significantly different across the tested conditions with regard to usefulness ($F(2,131) = 80.84, p= .000, \eta^{2}= .535$) and satisfaction ($F(2,131,) = 45.80, p= .000, \eta^{2}= .360$) of the concepts. Pairwise comparisons indicated that participants rated the interface as significantly useful compared to AVs without explicit communication interface. Additionally, the individual item comparisons in the usefulness dimension showed iHMI was significantly more effective compared to eHMI ($p=.015$). As for satisfaction items, no significant difference was found between eHMI and iHMI conditions ($p =1.000$).


\begin{figure}[ht]
\centering
\includegraphics[width = 0.9\linewidth]{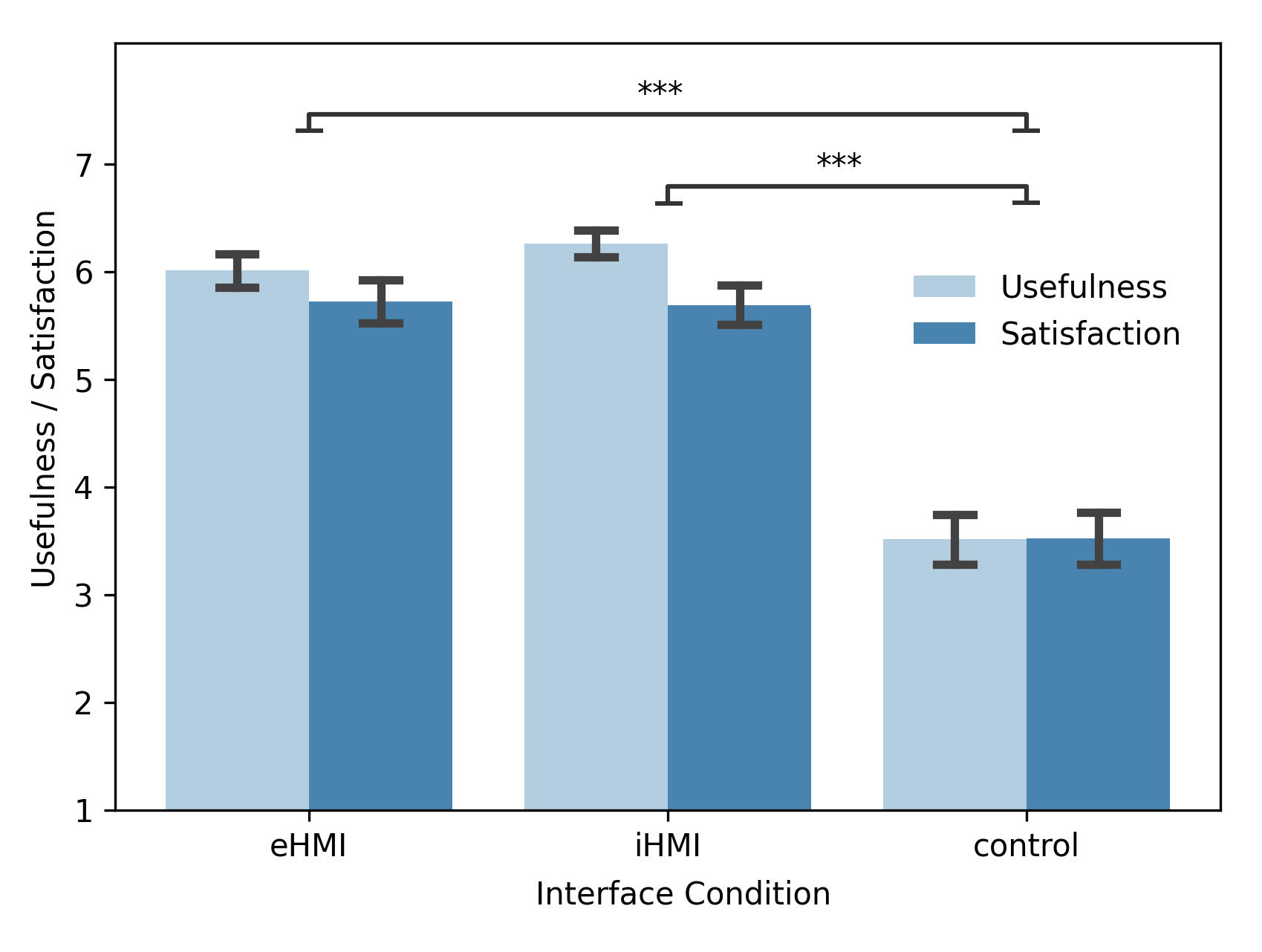}
\caption{Perceived usefulness and satisfaction of HMIs, where ‘***’ indicates $p < .001$. Note that the error bar showed the standard error.}
\label{fig:accept}
\end{figure}

\subsection{Pupil diameter and eye openness}
To understand how participants’ mental workload and vigilance \cite{chen2014pupilMW, martin2022pupillometry} were influenced by the interface condition, mean pupil diameter and eye openness changes were examined. Due to the eye tracker's technical issues, 6 participants' data were partially missing and were excluded from eye-tracking results. Therefore, eye-tracking measures were analyzed based on data from 38 participants.

\begin{figure}[ht]
\centering
\includegraphics[width = 0.9\linewidth]{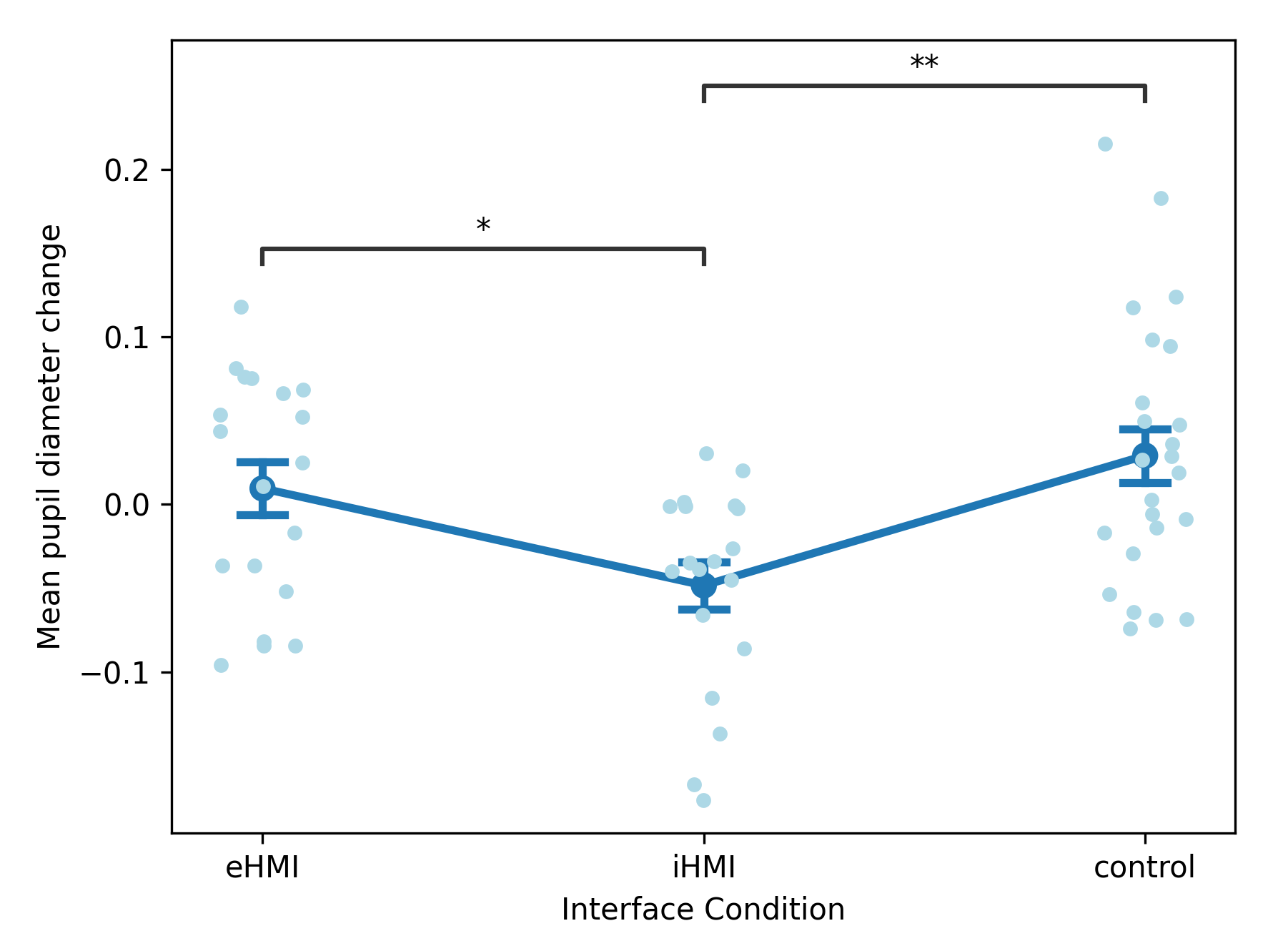}
\caption{Mean pupil diameter change and standard error across different interface conditions, where ‘*’ indicates $p < .05$,‘**’ indicates $p < .01$. Note that a positive change indicated larger pupil diameter following the interaction between CVs and AVs, while a negative change indicated smaller pupil diameter.} 
\label{fig:pupil}
\end{figure}
\begin{figure}[ht]
\centering
\includegraphics[width = 0.9\linewidth]{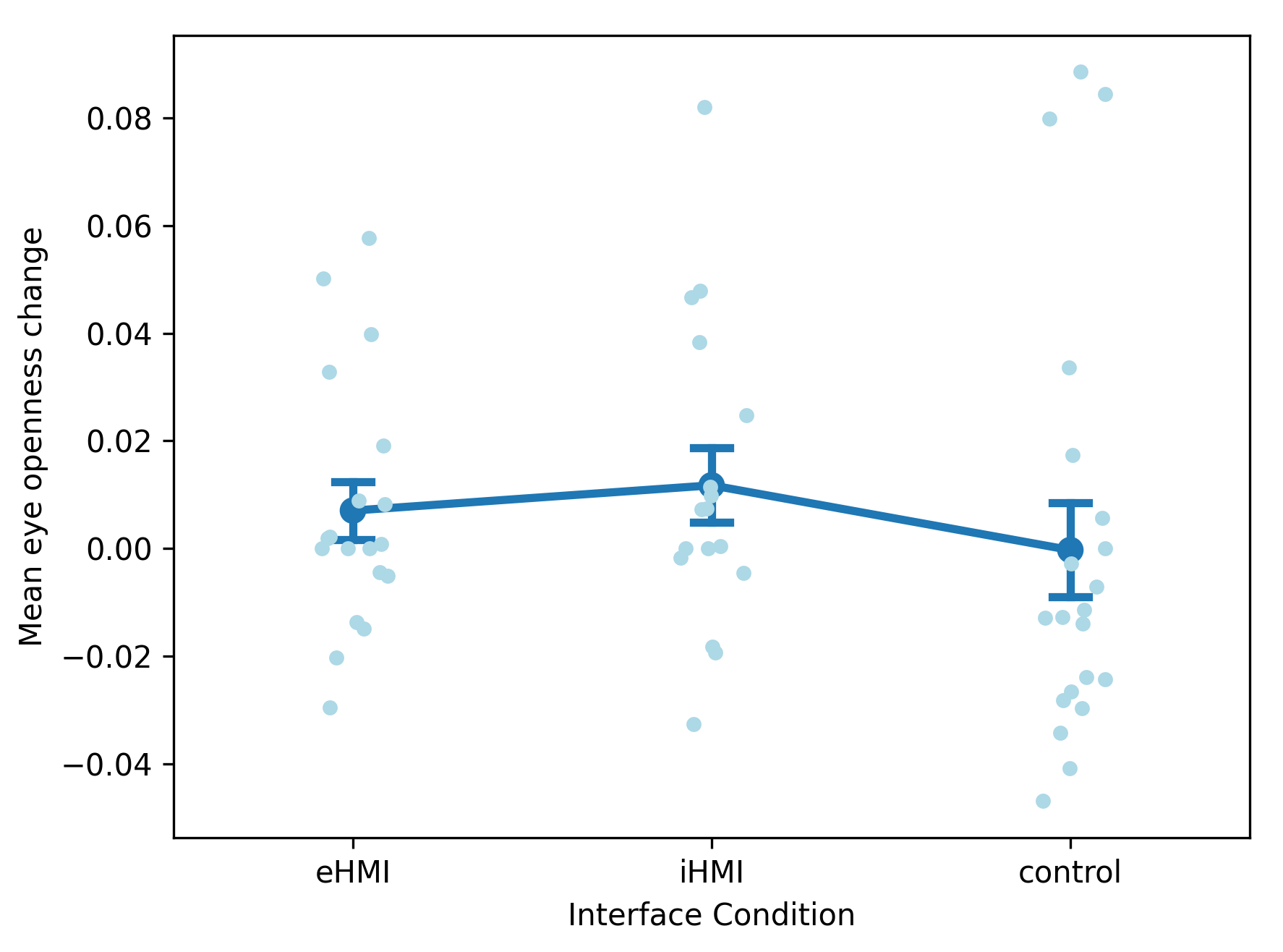}
\caption{Mean eye openness change and standard error across different interface conditions. Note that a positive change in eye openness indicated wider eyes following the interaction between CV drivers and AV, while a negative change indicated narrower eyes.}
\label{fig:openness}
\end{figure}

The results of one-way ANOVA performed on mean pupil diameter change showed that there was a significant difference among the three conditions ($F(2,113) = 6.69, p= .002, \eta^{2}= .185$). The post-hoc test showed that pupil diameter change in iHMI condition was significantly lower than that in the control ($p=.001$) and eHMI ($p=.042$) conditions (see Fig. \ref{fig:pupil}).
Regarding the mean eye openness change, the patterns showed that in the control condition, participants tended to frequently squeeze their eyes during the interaction compared to the eHMI and iHMI conditions (see Fig. \ref{fig:openness}). However, this difference was not statistically significant ($F(2,113) = 1.40, p= .253, \eta^{2}= .057$).

\subsection{CV speed}
\begin{figure}[ht]
\centering
\includegraphics[width = 0.9\linewidth]{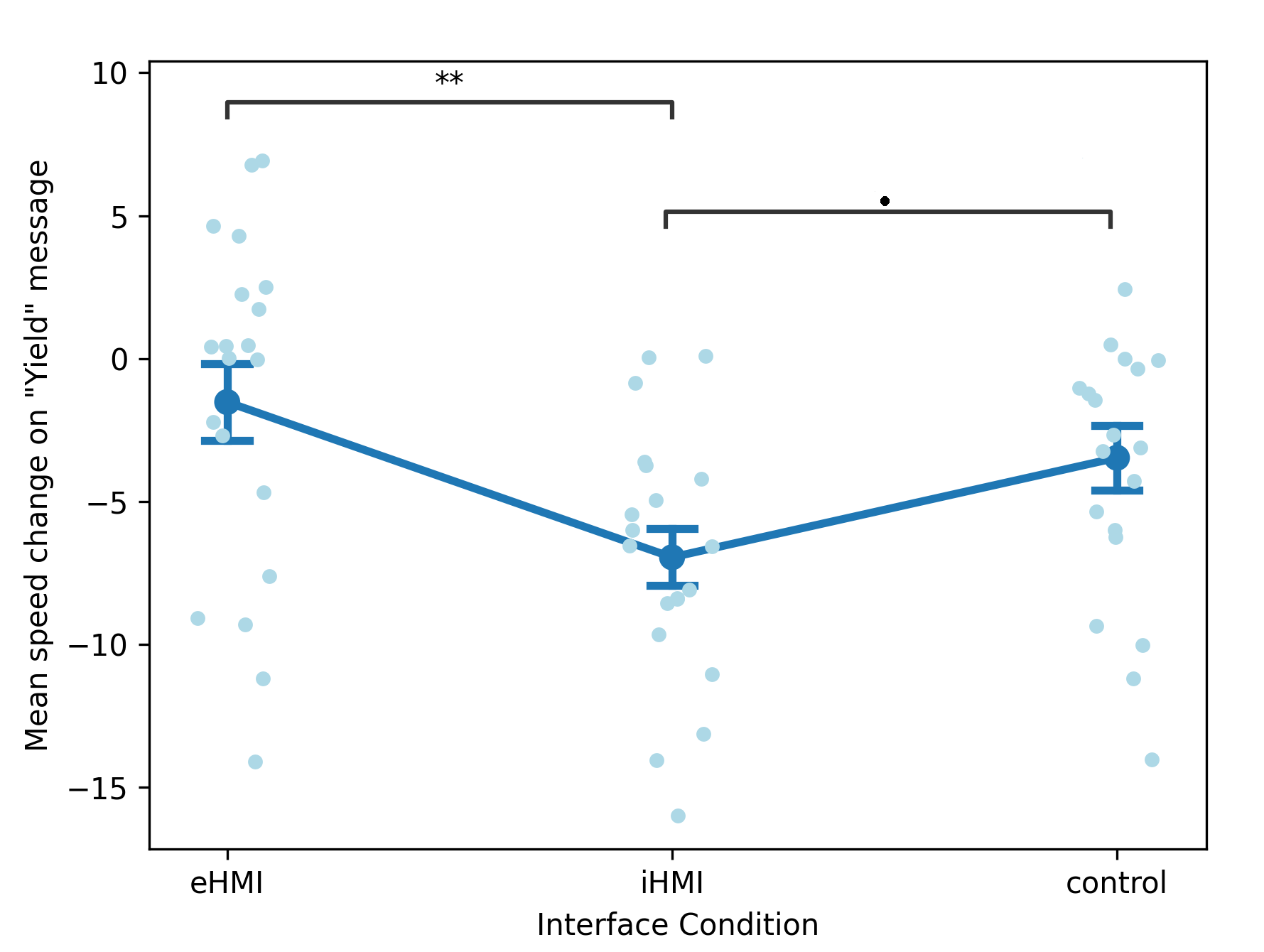}
\caption{Mean speed change and standard error across different interface conditions. The change was measured by calculating the average speed three seconds before and after receiving the message. Note that a positive change in speed indicated the CV had positive acceleration after onset of ``Yield'' message, while a negative change positive braking after the message.}
\label{fig:speed_change}
\end{figure}

\begin{figure}[ht]
\centering
\includegraphics[width = 0.9\linewidth]{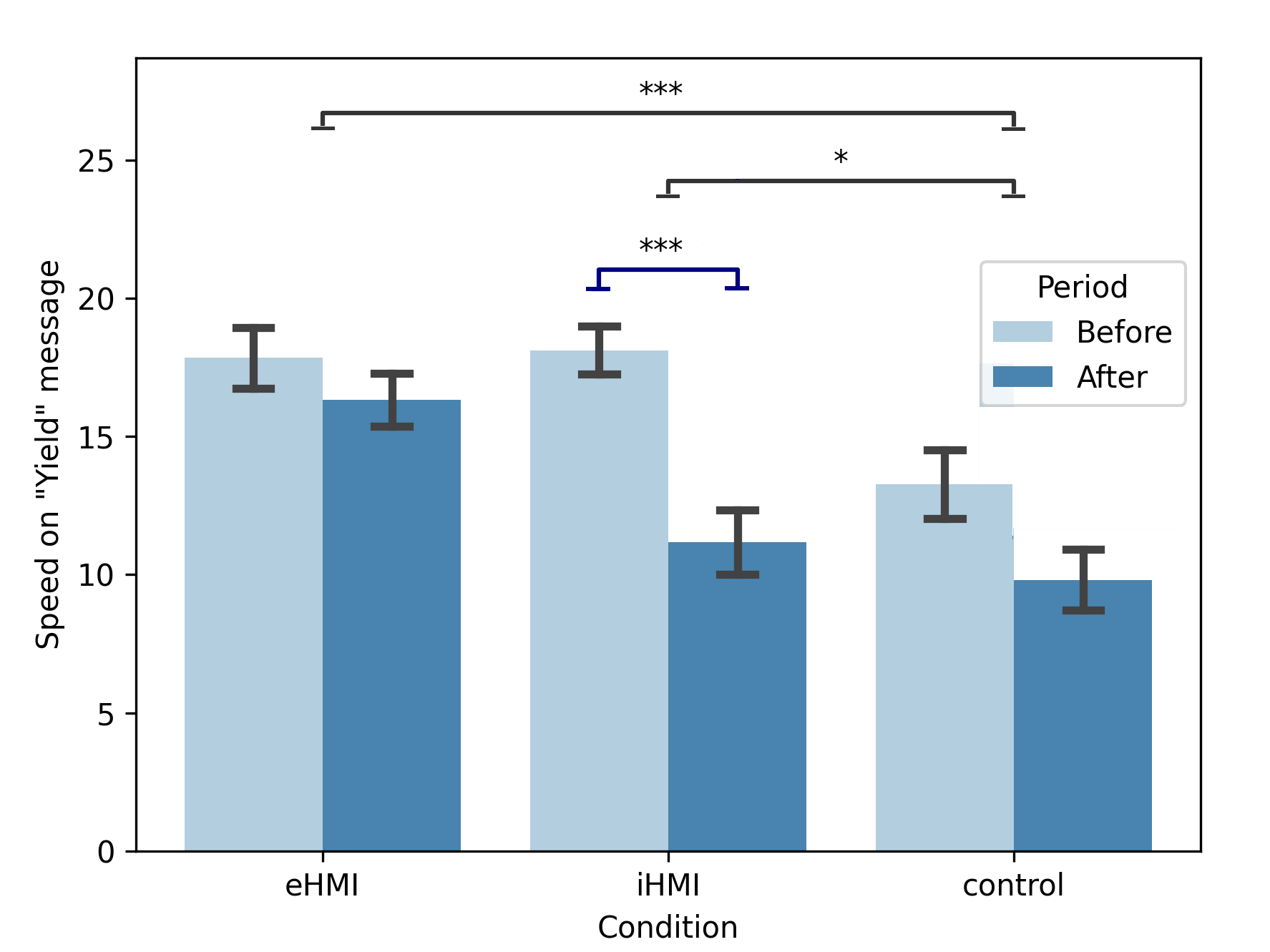}
\caption{Mean speed and standard error across different interface conditions measured in three second time period before and after onset of ``Yield'' message.}
\label{fig:speed_avg}
\end{figure}

The results of a one-way ANOVA conducted on the mean speed change revealed a significant difference among the three conditions ($F(2,119) = 5.59, p= .006$) when the AV insisted its right of way. According to the post-hoc test, the iHMI condition showed a significant drop in the CV speed after displaying the``Yield'' message compared to the eHMI condition ($p=.005$) and marginal drop compared to the control ($p=.096$) conditions (see Fig. \ref{fig:speed_change}). However, in the intersection where the AV yielded the right of way, the speed change was not statistically significant ($F(2,119) = 1.32, p= .548$). 
Comparing the average CV speed, the results showed that in control condition participants' speed was significantly lower than other conditions ($F(2,119) = 10.98, p= .000$). Specifically, at the intersection where the CV had the right of way to turn, the participants approached the intersection with significantly lower speed than that in the iHMI condition ($p=.005$). Similarly, at the ``Yield'' intersection, their speed was significantly lower than that in the iHMI ($p=.027$) and eHMI ($p=.000$) conditions (see Fig. \ref{fig:speed_avg}). 

\section{Discussions}

\subsection{Situation awareness}
In this study, the participants' SA level was measured in three interface conditions and the results indicated that the SA level was increased by either eHMI or iHMI (see Fig. \ref{fig:sa}). Compared to the control condition, in the eHMI and iHMI conditions, the participants were more conscious of the traffic situation. In particular, they were able to identify what the traffic issue was at the moment and who had the right of way to continue driving based on the shared information by giving quicker responses with respect to vehicle control. In the treatment conditions, immediately after receiving the visual message, the participants showed reactive behavior by slowing the car down and proceeded shortly after comprehending the message. In the control condition, drivers intended to keep a safe distance until they could understand the AV's intention based on the implicit cues of the AV (i.e., speed decrease or continuous driving). This finding was consistent with the research expectations and previous studies that HMIs filled the communication gap between AVs and CVs and reduced the ambiguity level in uncertain situations. 

\subsection{Trust}
The results regarding trust indicate that the designed HMIs had a significant positive impact on the participants' trust level (see Fig. \ref{fig:trust}). Irrespective of the interface conditions, participants exhibited higher trust in AVs compared to the control condition. Additionally, a notable behavioral difference was observed among the conditions. In the control condition, where the participants were informed that the AVs would not communicate with them, they tended to maintain a lower speed (see Fig. \ref{fig:speed_avg}) compared to the other conditions. However, their reaction to AVs was still delayed, as they took more time to decelerate than the AV required to cross the intersection.
As for the eHMI and iHMI conditions, the participants were more inclined to confidently rely on the HMI instructions in the iHMI condition, indicating a greater level of trust in the AV and quicker reaction, than in the eHMI condition.


\subsection{Acceptance}
With respect to the acceptance, the results showed that AVs were more acceptable with the designed HMIs than that in the control condition. Overall, both types of HMIs were highly acceptable compared to the control condition. However, the iHMI condition was more preferable, which might be due to the perceived ease of understanding the messages. First, the participants indicated that iHMI was grabbing their attention immediately and it was easy to see, unlike the eHMI for which they needed to be attentive to notice that AV was trying to communicate and put extra effort to see the message in the distance. Additionally, the participants' responses seemed to be faster in the iHMI condition which was reflected in the changes in vehicle speed when they received the message to yield the AV (see Fig. \ref{fig:speed_change}). Second, it was clear for them in the iHMI condition that the message was addressed to them, meaning that it was clear that the actions (i.e., yield or go) was displayed from the CV driver’s point of view and in a smaller distance, in contrast to the eHMI condition where the receiver could be other road users. This result was consistent with participants' ratings and with the previous studies \cite{eisma2021external} regarding to the effectiveness of interfaces, showing that the iHMI condition was more effective compared to the eHMI or control condition and resulted in faster responses.
\subsection{Eye tracking measures}

In this study, we utilized the pupil diameter and eye openness as metrics to evaluate participants' mental workload and vigilance. A significant difference was observed in the mean change of pupil diameter across the tested conditions, with the iHMI condition showing the greatest change compared to the control and eHMI conditions. The change in pupil diameter was established in previous studies as an indicator of changes in participants' vigilance and mental workload \cite{chen2014pupilMW, martin2022pupillometry}. Specifically, the pupil diameter tended to increase with an increase in mental effort and attention, which was often influenced by the level of uncertainty and perceived risk at the moment. As demonstrated in Fig. \ref{fig:pupil}, activation of the iHMI condition led to a noticeable decrease in pupil diameters, showing that the iHMI condition required less processing effort than the other tested conditions. 


As for the eye openness, we could not identify significant differences between the three conditions. One possible reason for this could be less sensitive as a measure to differentiate cognitive workload and vigilance compared to pupil diameters though it is a good indicator for driver fatigue or drowsiness \cite{zhou2022predicting, zhou2020driver}. Since each trial was conducted in a short period, (i.e., 5 minutes), it might not be sufficient to impact participants' vigilance. A longer experiment duration could potentially provide more reliable results by giving participants more time to adapt to the experimental conditions. 

\subsection{Implications}
Effective communication between vehicles is a vital aspect to develop collaborative driving in a mixed traffic environment. However, due to the complexity of understanding the capabilities and intentions of AVs, it has become more challenging for drivers in CVs to maintain the necessary level of SA, which is critical to ensure transportation safety in mixed traffic. This study showed that SA was improved during the AV-CV interaction using external or internal HMIs at an intersection. Results indicated that sharing the traffic situation and the AV's intention via appropriate designed HMIs boosted human-drivers' SA. However, there was no standardized communication protocols so far, which might lead to confusion and misinterpretation of signals. Our study provided an example of designing HMIs with a human-centered process, and showed that iHMIs might be more advantageous than eHMIs. More research should be conducted to understand the potential of iHMIs and provide more standard communication protocols, involving design (e.g., visual cues, auditory signals), location of HMIs, and other elements in the communication protocols. 

Moreover, as self-driving technology evolves, establishing public acceptance and trust becomes crucial. Ambiguity in communication between CVs and AVs can raise concerns among road users about the reliability and safety of autonomous systems. Our findings showed that the proposed HMI designs increased their trust in AVs compared to the control condition. We should also point out that extra information in ambiguous scenarios could potentially add additional workload on driving tasks. Therefore, it is important to consider that when designing HMIs for vehicle communication systems to support real-time decision making. Overall, our findings offer valuable insights for future investigations and implementations in AV-CV communication investigation in mixed traffic.

\subsection{Limitations}
This research also has imitations that can be addressed in the future studies. First, more objective measures (e.g., eye fixation on areas on interests) should be collected to better understand attention requirements of each design. Due to the limitations of the VR headset, we were not abel to collect such measures. Second, only one particular scenario was investigated to understand the effects of the proposed HMIs. In future studies, more scenarios should be included to generalize the results for various ambiguous traffic situations, and more road users (e.g., pedestrians and other vehicles) should be involved to make the scenarios close to the real traffic situations. Third, the study population mainly included university students and was not gender balanced. Future studies should include a more diverse sample to better understanding the effects of HMIs on AV-CV communications in mixed traffic. 

\section{Conclusions}

In this research, we aimed to develop HMIs that facilitate communication between AV and CV drivers and investigate how such interfaces would influence CV drivers’ SA, trust, acceptance, cognitive workload, and vigilance when navigating in mixed traffic environments, particularly at intersections. We designed eight different interface concepts and subsequently tested the highest-rated concepts:internal and external HMIs.  We also included a control condition for comparison. The effectiveness of the HMIs was evaluated using SA, trust, acceptance, cognitive workload, and vigilance using participants' self-reporeted and eye-tracking measures in ambiguous situations where the CV needed to make a left turn at an intersection with malfunctioning traffic lights. We found that HMIs were assisting CV drivers in uncertain situations and resulted in increase of SA level, trust, and acceptance. The iHMI was considered the most effective communication method with AV and resulted in lowest change in drivers' mental workload.


\bibliography{main}
\begin{IEEEbiography}[{\includegraphics[width=1in,height=1.55in,clip,keepaspectratio]{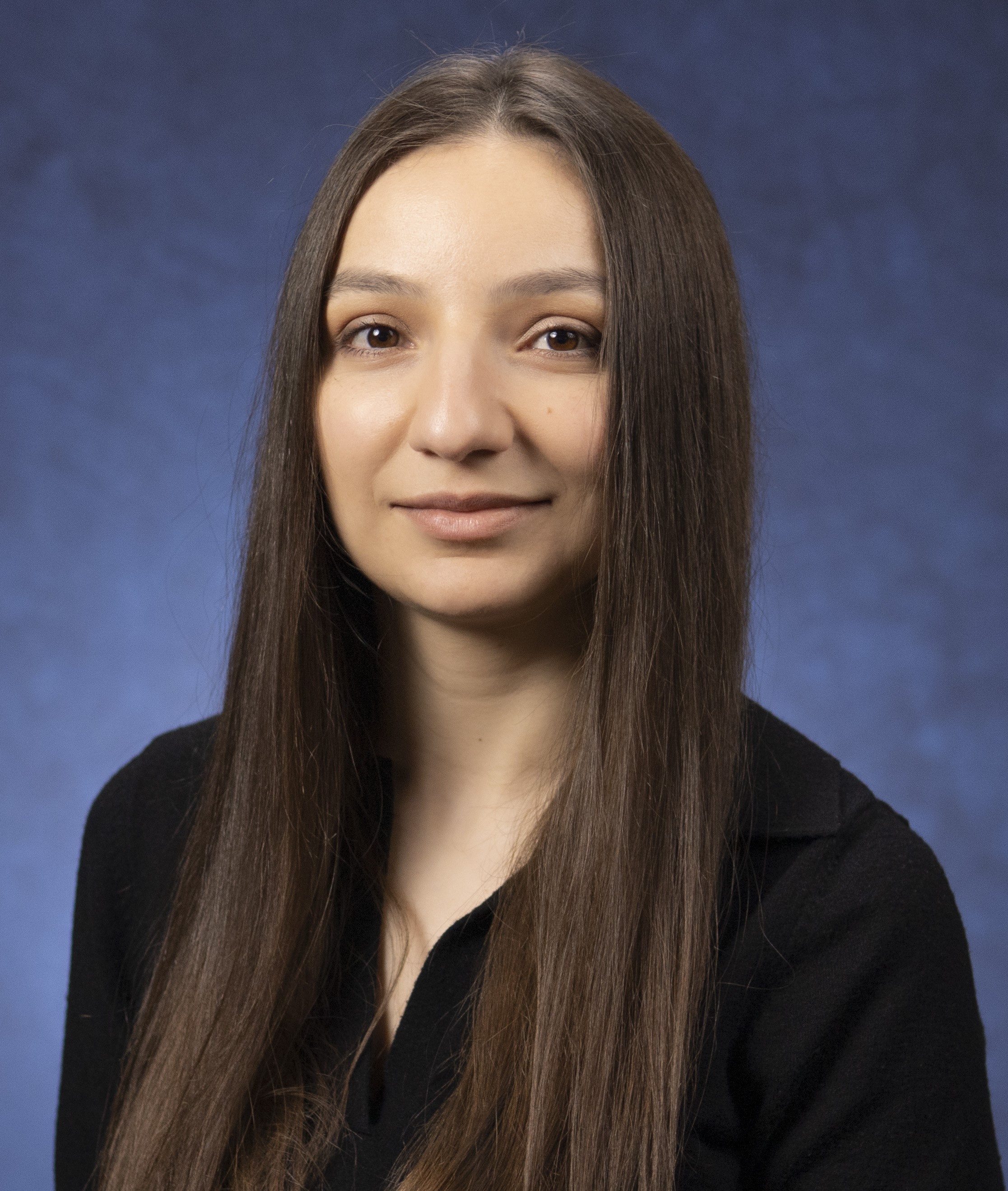}}]{Lilit Avetisyan received her B.E. degree in 2017 and MS degree in 2019 in Information Security from the National Polytechnic University of Armenia. She is currently pursuing her Ph.D. degree in Industrial and Systems Engineering at the University of Michigan, Dearborn. Her main research interests include human-computer interaction, explainable artificial intelligence and human-centered design.}
\end{IEEEbiography}

\begin{IEEEbiography}[{\includegraphics[width=1in,height=1.55in,clip,keepaspectratio]{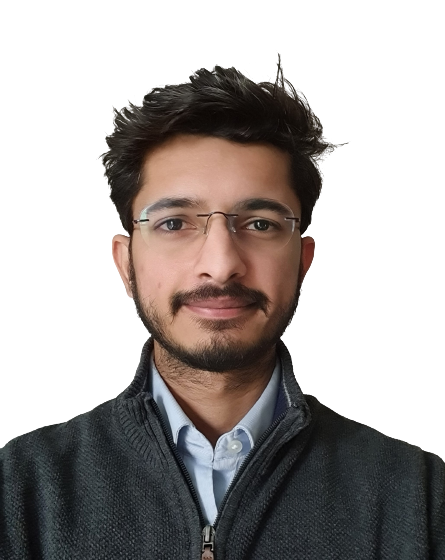}}]{Aditya Deshmukh received his B.E. degree in 2017 in Electronic and Telecommunication. He is currently pursuing M.S. in Human-Centered Design and Engineering at the University of Michigan, Dearborn. His main research interests include human-computer interaction, human-AI interaction, and human-centered design.}
\end{IEEEbiography}

\begin{IEEEbiography}[{\includegraphics[width=1in,height=1.55in,clip,keepaspectratio]{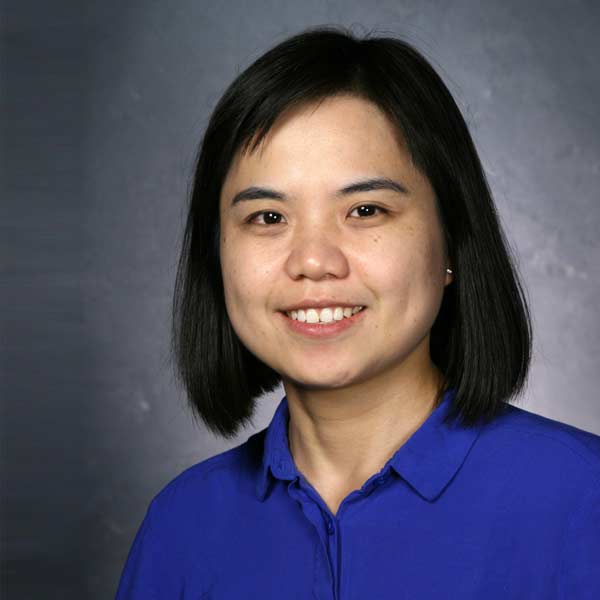}}]{X. Jessie Yang is an Assistant Professor in the Department of Industrial and Operations Engineering, University of Michigan, Ann Arbor. She earned a PhD in Mechanical and Aerospace Engineering (Human Factors) from Nanyang Technological University, Singapore. Dr. Yang’s research include human-autonomy interaction, human factors in high-risk industries and user experience design.}
\end{IEEEbiography}

\vspace{-20pt}

\begin{IEEEbiography}[{\includegraphics[width=1in,height=1.55in,clip,keepaspectratio]{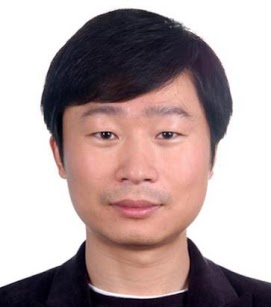}}]{Dr. Feng Zhou received the Ph.D. degree in Human Factors Engineering from Nanyang Technological University, Singapore, in 2011 and Ph.D. degree in Mechanical Engineering from Gatech Tech in 2014. He was a Research Scientist at MediaScience, Austin TX, from 2015 to 2017. He is currently an Assistant Professor with the Department of Industrial and Manufacturing Systems Engineering, University of Michigan, Dearborn. His main research interests include human factors, human-computer interaction, engineering design, and human-centered design.}
\end{IEEEbiography}
\end{document}